\documentclass[default]{sn-jnl}

\theoremstyle{thmstyleone}%
\theoremstyle{thmstyletwo}%

\theoremstyle{thmstylethree}%

\begin{document}

\title[Detachment of a rigid flat punch from a viscoelastic material]{Detachment of a rigid flat punch from a viscoelastic material}


\author[1,2]{\fnm{Antonio} \sur{Papangelo}}\email{antonio.papangelo@poliba.it}

\author*[1,2]{\fnm{Michele} \sur{Ciavarella}}\email{mciava@poliba.it}


\affil[1]{\orgdiv{Department of Mechanics Mathematics and Management}, \orgname{Politecnico di Bari}, \orgaddress{\street{via Orabona 4}, \city{Bari}, \postcode{70125}, \state{Italy}}}

\affil[2]{\orgdiv{Department of Mechanical Engineering}, \orgname{Hamburg University of Technology}, \orgaddress{\street{Am Schwarzenberg-Campus 1}, \city{Hamburg}, \postcode{21073}, \state{Germany}}}


\abstract{We show that the detachment of a flat punch from a viscoelastic substrate has
a relatively simple behaviour, framed between the Kendall's elastic solution at
the relaxed modulus and at the instantaneous modulus, and the cohesive
strength limit. We find hardly any dependence of the pull-off force on the
details of the loading process, including maximum indentation at preload and
loading rate, resulting much simpler than the case of a spherical punch.
Pull-off force peaks at the highest speeds of unloading, when energy
dissipation is negligible, which seems to be in contrast with what suggested
by the theories originated by de Gennes of viscoelastic semi-infinite crack
propagation which associated enhanced work of adhesion to dissipation.}




\maketitle
\section{Introduction}

Over the last two decades, there has been a tremendous interest in soft
adhesive interfaces in various areas of technology, and in particular, those
based on the van der Waals forces, which have been largely inspired by nature.
Bio-inspired technologies are a growing area in robotics and
grasping/pick-and-place manipulation \cite{Gio2021a}\cite{Mazzolai2019}, since
they show several advantages such as high adhesive strength competitive to
suction devices, more sustainable technology, recyclability, absence of
residues \cite{Gio2021b}, adaptability \cite{VCacu} and less energy
consumption with respect to classical solutions \cite{Shui}. Examples of
application in several emerging technologies are described in \cite{Arzt}, in
robotics \cite{Lunni2020}\cite{Lunni2018} and in pick-and-place manipulation,
also in space \cite{Asbeck,Shintake}.

In hard materials, van der Waals adhesive forces are destroyed very easily by
roughness, and Dahlquist postulated, based on experiments on Pressure
Sensitive Adhesives (PSA), that the elastic modulus for them should be less
than about $0.3$ MPa \cite{Dahlquist} for sufficient adhesion. Soft materials
are thus widely adopted in engineering, and those are often viscoelastic,
which is pushing the research in the area of viscoelastic crack growth in soft
media like polymers or biological materials \cite{Creton,CPM2021}. Early
extensive measurements mostly with peeling setups \cite{Barquins1981,GentSchultz,Gent,Andrews,Greenwood1981,Maugis}
found that the peeling load increased very significantly with peeling speed.
Since work of adhesion is equal to load for an elastic peeling problem
according to the {Rivlin classical solution for the typical $90^\circ$ peeling conditions} \cite{Rivlin} for which the force
per unit lenght of elastic peeling is exactly the work of adhesion, measuring
the force in a viscoelastic tape generated the concept of \textquotedblleft
apparent work of adhesion $\Delta\gamma$" obtained as the product of the adiabatic
value $\Delta\gamma_{0}$ and a function of crack velocity $V$ of the contact/crack
line and temperature, often in the form of a power law%
\begin{equation}
\frac{\Delta\gamma\left(  V\right)  }{\Delta\gamma_{0}}=1+\left(  \frac{V}{V_{ref}%
}\right)  ^{n}\label{wvisco}%
\end{equation}
where $V_{ref}=\left(  ka_{T}^{n}\right)  ^{-1}$ and $k,n$ are constants with
$0<n<1$ and $a_{T}$ is the {Williams-Landel-Ferry (WLF) shift factor} to translate viscoelastic modulus
results at various temperatures $T$ \cite{Williams}. 
{The Williams-Landel-Ferry equation (or WLF equation) permits time-temperature superposition by the following formula}
{
\begin{equation}
\log _{10}a_{T}=-\frac{C_{1}\left( T-T_{r}\right) }{C_{2}+\left(
T-T_{r}\right) }
\end{equation}
}
{where $T$ is
the temperature, $T_{r}$ is a reference temperature chosen to
construct the compliance master curve and $C_{1},C_{2}$ are
empirical constants adjusted to fit the values of the superposition
parameter $a_{T}$. In other words, measurements at different
temperatures permit to obtain a wide range of frequency estimate of the
complex modulus of viscoelastic material which would otherwise not be
obtained in a single experiment with a restricted range of frequencies via
\begin{equation}
E\left( \omega ,T\right) =E\left( a_{T}\omega ,T_{r}\right) 
\end{equation}}

{Equation (\ref{wvisco}) is known}
as the Gent-Schultz \textquotedblleft empirical law" \cite{GentSchultz}, and
is widely adopted even to different geometries, although the constants may
depend on geometry (in peeling itself, there is certainly an effect of the
angle of peeling). Notice that in this form there is no indication about the
maximum enhancement, although the connection to the WLF factor indicates a
link with elastic modulus whose increase has clear limits. However, peeling
experiments such as Gent and Petrich \cite{gentpetrich} showed a maximum in
peel force followed by a decrease in an unstable regime and stick-slip. {It is remarkable that peeling experiments are still not entirely understood: in a peeling geometry of a viscoelastic tape it would seem that speed-dependence of the load of typical $90^\circ$ angles would be precluded in a rate-independent cohesive stress model (see a detailed recent study in \cite{Ceglie}) like the present model or the Knauss-Schapery or de Gennes-Persson-Brener theories we are discussing in the present paper, yet it is clearly observed. The case of viscoelastic peeling seems to require rate-dependence in the cohesive (or the cut-off) stress.}

From a more fundamental perspective, the need of Cohesive Zone Model (CZM)
formulations to describe the crack propagation in viscoelastic media was
evident in the 1970's \cite{Rice,Graham,Schapery,SchaperyII,Knauss}. Knauss-Schapery showed that when the stress field is well
defined by an \textquotedblleft applied" stress intensity factor $K_{A}$, the
speed of propagation was defined by a generalized Irwin equation which for
pure mode I reads%
\begin{equation}
\Delta\gamma_{0}=C\left(  t_{b}\right)  K_{A}^{2} \label{schap1}%
\end{equation}
where the linear viscoelastic creep compliance $C\left(  t\right)  $ replaces
the elastic compliance ($\left(  1-\nu^{2}\right)  /E$ in plane strain, $1/E$
in plane stress, where $E$ is Young's modulus and $\nu$ is Poisson's ratio) in
the very similar Linear Elastic Fracture Mechanics (LEFM) critical equivalent
condition, and $t_{b}$ is an effective time of relaxation in the cohesive
zone. This equation is valid for any geometry, provided the cohesive zone
length $b$ is small with respect to other length scales in the problem. In
other words, in this model the "true" work of adhesion is rate-independent,
and yet it results in load enhancement.

An alternative approach returns to the concept of the "apparent work of
adhesion" and stems from the seminal work by de Gennes \cite{deGennes} who
attributed it to dissipation possibly very far ahead from the crack tip in the
so-called "liquid" zone --- a concept which Persson and Brener
\cite{Persson2005} elaborated in more quantitative terms, still for
semi-infinite cracks. Despite the different approaches, both dissipation-based
and CZM-based theories provide similar results for semi-infinite cracks and
obtain the same maximum toughness enhancement \cite{hui}
\begin{equation}
\frac{\Delta\gamma\left(  \infty\right)  }{\Delta\gamma_{0}}=\frac{E_{\infty}}{E_{0}%
}\label{limit-enhancement}%
\end{equation}
being $E_{0}$ and $E_{\infty}$ the so-called relaxed and instantaneous modulus
of the viscoelastic material, if we interpret as "apparent work of adhesion"
\begin{equation}
\Delta\gamma\left(  V\right)  =\frac{K_{A}^{2}}{E_{0}}%
\end{equation}

The question is less clear for any finite size body \cite{CiaPap2021,CiaPapMec2022}, for which de Gennes postulated dissipation is restricted
in space and so should reach a maximum and then decay with speed
\cite{deGennes}. This was intended to perhaps explain the well known Gent and
Petrich \cite{gentpetrich} results for peeling experiments which we have
mentioned. {However, while in peeling experiments load and work
of adhesion seem to be identical (at least in elastic case), 
this cannot be generalized to generate the confusion in the literature between ”load” enhancement and ”work of adhesion” enhancement, in turn apparently due to viscoelastic dissipation.} Is increase of load necessarily linked to
viscoelastic dissipation? Certainly this is not evident from the Irwin
equation of Schapery (Eq. (\ref{schap1})), which in the limit of very low or very
fast speed gives rubbery or glassy \textit{elastic }materials, and yet
provides the enhancement.

For example, in \cite{xuHui} the authors study a double cantilever beam (DCB)
specimen under a vertical force at constant distance from the crack tip
(recalling a linearized version of the peeling problem) finding the maximum
dissipation at intermediate crack propagation speed, but a monotonic increase
of the load with speed. Viceversa, in the case where the distance is permitted
to increase without limit, like when applying a concentrated moment to a DCB
in a fixed point while the crack travels, dissipation seems to increase with
load like in de Gennes' theory  \cite{CiaPapMec2022}.  In the case of
indentation of a halfspace by a punch, we may expect that the finite radius of
the punch may play a role. Indeed, while stresses in a semi-infinite crack
decay with distance from crack tip as $r^{-1/2}$, those far from the punch in
the infinite halfspace decay like $r^{-2}$ from the resultant force hence much
faster. So, although at high speed of crack propagation there must be still a
"liquid" region of transition from glassy to rubbery state, this region may
dissipate less than the analogous case of the semi-infinite crack. {It should be mentioned that Knauss \cite{Knauss} tested and studied
theoretically already some of the effects of finite thickness in the pure
shear geometry test, and also did not find a peak in load with crack speeds,
which are not really explained by the dissipation limitations at the basis
of de Gennes and Persson's finite size theories.} We shall
therefore investigate in details how load-enhancement varies with speed in a
finite size contact, and if this is related to viscoelastic dissipation.

Recently, a few studies have been devoted to the Hertzian geometry finding
complex dependences on preload and loading rate (and not only unloading rate)
\cite{ciava2021,AffVio2022,VioAff2022size,VioAffRange,MuserPer}. Surprisingly,
less attention has been paid to the apparently simpler problem of the flat
punch geometry, {apparently also because in practise it requires special care for
avoiding misalignments in an experimental setup.}
Here, by adopting the Boundary Element Method (BEM)\ we
therefore explore numerically this problem with the support of extension of
classical theories for the elastic case derived by Maugis \cite{Maugis} (see
also \cite{TangHui}) to the viscoelastic case.

\section{The mechanical model}

\begin{figure}[t]
\begin{center}
\includegraphics[width=4.3799in]{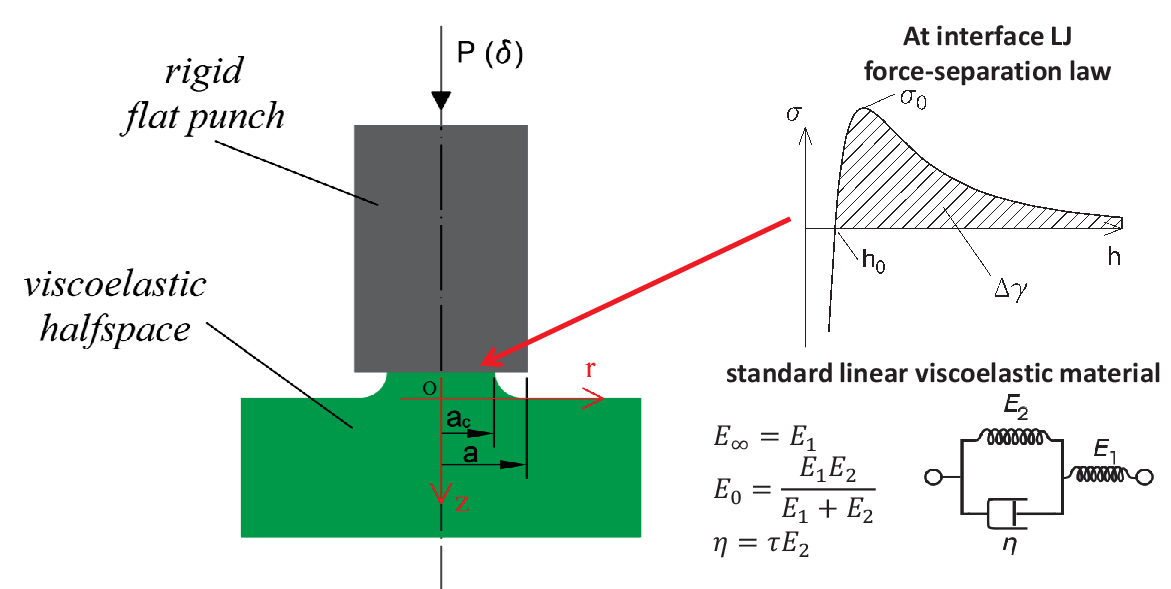}
\end{center}
\caption{Schematic representation of the problem considered. An axisymmetric
flat punch of radius $a$ indents a linear viscoleastic halfspace. A standard
material is assumed with two elastic moduli $E_{0}$ and $E_{\infty}$ and a
single relaxation time $\tau$. The vertical displacement of the half-space
$u_{z}$ and the indentation $\delta$ are positive when the punch indents the
halfspce. $a_{c}$ is the radius of the ligament of the crack, and is time
dependent. The punch and the substrate interact through a Lennard-Jones (LJ)
force separation law.}%
\label{Figschema}%
\end{figure}

The adhesive behaviour of an axisymmetric rigid flat punch of radius $a$ in
frictionless contact with a viscoelastic half-space is studied. For the linear
viscoelastic material, the standard model is assumed, constituted by a spring
placed in series with an element constituted by a dashpot and a spring in
parallel (Fig. \ref{Figschema}). The viscoelastic material has the relaxed
Young's modulus $E_{0}$,  instantaneous modulus $E_{\infty}\footnote{This
notation stems from frequency dependence notation of the complex modulus
$E\left(  \omega\right)  $ so that $E_{0}=E\left(  0\right)  $ is the relaxed
modulus. Other authors use the notation that $E_{0}$ is the instantaneous
modulus as obtained at time $t=0$ in a relaxation test.}$, and the relaxation
time $\tau$. The flat punch interacts with the viscoelastic substrate via a
Lennard-Jones 3-9 force-separation law%
\begin{equation}
\sigma\left(  h\right)  =\frac{8\Delta\gamma}{3h_{0}}\left[  \left(
\frac{h_{0}}{h}\right)  ^{3}-\left(  \frac{h_{0}}{h}\right)  ^{9}\right]
\label{LJ}%
\end{equation}
where $\sigma$ is the traction ($\sigma>0,$ when tensile), $h$ is the gap and
$h_0$ the equilibrium distance (the maximum tensile stress $\sigma
_0=\frac{16\Delta\gamma}{9\sqrt{3}h_0}$ takes place at separation
$h=3^1/6h_0$). The gap function is written as%
\begin{equation}
h(r)=-\delta+h_{0}+u_{z}\left(  r\right)  \label{h}%
\end{equation}
where $r$ is the radial coordinate, $\delta>0$ when the flat punch approaches
the viscoelastic half-space, $u_z\left(  r\right)  $ is the deflection of the
viscoelastic half-space, which will depend on the loading history.

For elastic axisymmetric problems \cite{Greenwood1981,Feng}%

\begin{equation}
u_{z}\left(  r\right)  =-\frac{1}{E^{\ast}}%
{\displaystyle\int}
\sigma\left(  s\right)  G\left(  r,s\right)  sds \label{int}%
\end{equation}
where $1/E^{\ast}=\left(  \left(  1-\nu_{1}^{2}\right)  /E_{1}+\left(
1-\nu_{2}^{2}\right)  /E_{2}\right)  $ is the composite elastic modulus of the
two bodies in contact, $\sigma\left(  s\right)  $ is the pressure distribution
(the minus sign account for the fact that we considered compressive tractions
as negative), $G\left(  r,s\right)  $ is the kernel function%

\begin{equation}
G\left(  r,s\right)  =\left\{
\begin{array}
[c]{cc}%
\frac{4}{\pi r}K\left(  \frac{s}{r}\right)  , & \qquad\qquad s<r\\
\frac{4}{\pi s}K\left(  \frac{r}{s}\right)  , & \qquad\qquad s>r
\end{array}
\right.  \label{kernel}%
\end{equation}
and $K\left(  \boldsymbol{k}\right)  $ is the complete elliptic integral of
the first kind of modulus $\boldsymbol{k}$. Hence, by using the
elastic-viscoelastic correspondence principle in the form of Boltzmann
integrals \cite{Chri}, the normal displacements of the viscoelastic half-space
$u_{z}(r,t)$ at time $t$, at location $r$ depend on the pressure history as%

\begin{equation}
{u_{z}(r,t)=\frac{-1}{E_{0}^{\ast}}\int\int_{-\infty}^{t}c(t-\tau
)\frac{d\sigma(s,\tau)}{d\tau}G\left(  r,s\right)  sd\tau ds} \label{int2}%
\end{equation}
where ${c(t)}$ is the dimensionless creep compliance function, which for a
standard linear viscoelastic solid is%

\begin{equation}
c(t)=E_{0}^{\ast}C\left(  t\right)  =\left[  1+\left(  k-1\right)  \exp\left(
-\frac{t}{\tau}\right)  \right]
\end{equation}
where $k=E_{0}/E_{\infty}$.

Equation (\ref{int2}), discretized in time and space, is solved to provide the
adhesive solution sought. The numerical scheme is alike that in Ref.
\cite{Greenwood1981,Feng,PapCia2020} except for accounting the viscoelastic
behavior of the half-space. The radial domain is discretized with $N$ equally
spaced elements, so that we have $M=N+1$ nodes where Eq. (\ref{int2}) is
solved. From one node to the other a linear variation of the normal traction
$\sigma\left(  s\right)  $ is assumed, which is usually referred as "the
method of the overlapping triangles" \cite{Joh}, and a sequential continuation
algorithm is adopted in time, so that the solution $u_{z}\left(
r,t_{i}\right)  $ at time $t_{i}$ is used as a guess for the next time step
$t_{i+1}=t_{i}+\Delta t$, where the time increment $\Delta t$ is kept fixed
during the simulations. More details of the numerical implementation have been
given in \cite{PapCia2020}.

\section{From LEFM to uniform debonding behaviour}

The contact of a rigid flat punch can be considered as an external crack that
propagates at the interface under the action of a tensile load. If one assumes
the punch to be rigid and the substrate \textit{elastic}, neglecting friction,
Maugis \cite{Maugis} shows%

\begin{equation}
G=\frac{K_{I}^{2}}{2E^{\ast}}=\frac{P^{2}}{8\pi a^{3}E^{\ast}}%
\end{equation}
where $G$ is the energy release rate, $P$ is the normal load and
$K_{I}=P/\sqrt{4\pi a^{3}}$ is the mode-I stress intensity factor. Imposing
the Griffith energy balance $G=\Delta\gamma$ \cite{Kendall1971,Maugis}
one finds the peeling force to be%

\begin{equation}
P=\sqrt{8\pi E^{\ast}\Delta\gamma a^{3}}%
\end{equation}
where $E^{\ast}=E/\left(  1-\nu^{2}\right)  .$ Correspondingly the pull-off
stress $\widehat{\sigma}_{po}=\max\left(  \widehat{\sigma}\right)  $ is%
\begin{equation}
\sigma_{po}=\sqrt{\frac{8E^{\ast}\Delta\gamma}{\pi a}}%
\end{equation}
which scales with the punch radius as $a^{-1/2}$, a well-known result of
Linear Elastic Fracture Mechanics (LEFM). Clearly, $\sigma_{po}$ cannot be
larger then the theoretical strength of the interface $\sigma_{0}$. Hence, we
define a lengthscale%

\begin{equation}
a_{0}=\frac{8E^{\ast}\Delta\gamma}{\pi\sigma_{0}^{2}}%
\end{equation}
and only for a punch of radius $a\gg a_{0}$ one expects LEFM based model to be
accurate \cite{Kendall1971}, while for $a\ll a_{0}$ the cohesive limit is
asymptotically obtained, where $\sigma_{po}\rightarrow\sigma_{0}$.

If the half-space is viscoelastic, we can consider it to behave as an elastic
half-space in the limit of very slow and very fast unloading rate. In the
latter case one obtains two different lengthscales $\left\{  a_{0}%
,a_{1}\right\}  $ which refer respectively to the slow and fast unloading
limits (see also Ciavarella (\cite{ciava2022}) for the analogous case of
"short cracks"). Recalling $k=E_{0}/E_{\infty}<1$, and using $\widehat{a}%
_{0}=a_{0}/h_{0}$ and $\widehat{\sigma}_{po}=\sigma_{po}/\sigma_{0}$ one gets
in dimensionless form%

\begin{align}
&  \left\{
\begin{array}
[c]{l}%
\widehat{\sigma}_{po}=\sqrt{\frac{9\sqrt{3}}{2\pi\widehat{a}\Sigma_{0}}}<1\\
\widehat{a}_{0}=\frac{9\sqrt{3}}{2\pi\Sigma_{0}}%
\end{array}
\right.  ,\qquad\qquad\text{slow limit}\label{slowlimit}\\
&  \left\{
\begin{array}
[c]{l}%
\widehat{\sigma}_{po}=\sqrt{\frac{9\sqrt{3}}{2\pi\widehat{a}\Sigma_{0}k}}<1\\
\widehat{a}_{1}=\frac{9\sqrt{3}}{2\pi k\Sigma_{0}}%
\end{array}
\right.  ,\qquad\qquad\text{fast limit} \label{fastlimit}%
\end{align}
where $\Sigma_{0}=\sigma_{0}/E_{0}^{\ast}$.

\section{Numerical results}

\begin{figure}[t]
\begin{center}
\includegraphics[width=4.3799in]{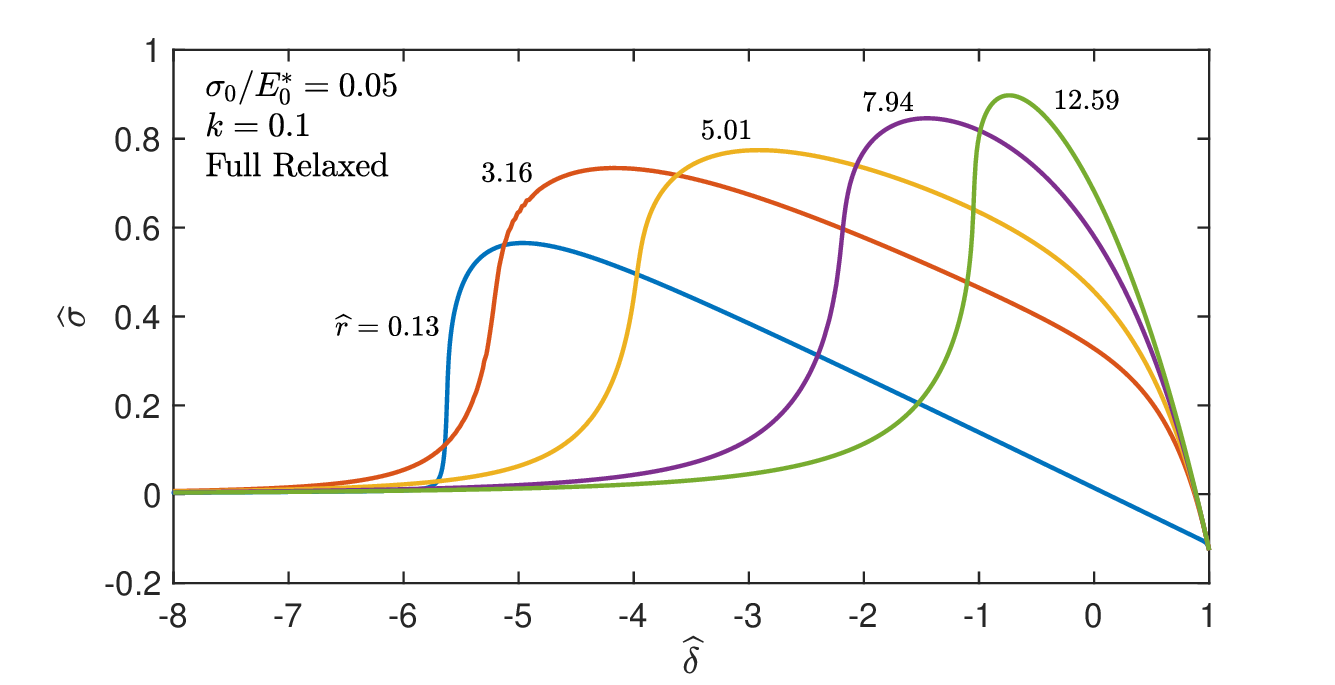}
\end{center}
\caption{Unloading curves for $\Sigma_{0}=0.05$, $k=0.1$, punch of radius
$\widehat{a}/\widehat{a}_{0}=2.02$,\ unloaded from a fully relaxed
viscoelastic substrate starting from $\widehat{\delta}_{0}=1$ at different
unloading rates $\widehat{r}=\left[  0.13,3.16,5.01,7.94,12.59\right]  $.}%
\label{FigUnload}%
\end{figure}

We conducted a wide campaign of numerical investigations, whose results are
reported in this section in dimensionless form, so that the following
quantities have been defined%

\begin{equation}
\widehat{a}=\frac{a}{h_{0}};\qquad\widehat{\sigma}=\frac{\sigma}{\sigma_{0}%
};\qquad\widehat{P}=\frac{P}{\pi a^{2}\sigma_{0}};\qquad\widehat{\delta}%
=\frac{\delta}{h_{0}};\qquad\widehat{t}=\frac{t}{\tau};
\end{equation}
while $\widehat{\sigma}_{po}=\max\left(  \widehat{\sigma}\right)  $ is the
pull-off stress. Unless differently stated, we used in our simulations
$N=200,$ $\Sigma_{0}=0.05$ and $k=0.1$. First, let us consider the case of a
punch of radius $\widehat{a}/\widehat{a}_{0}=2.02$ that is unloaded from a
fully relaxed viscoelastic substrate starting from $\widehat{\delta}%
_{0}=\widehat{\delta}\left(  \widehat{t}_{0}=0\right)  =1$ at different
unloading rates $\widehat{r}=r\tau/h_{0}=\left[
0.13,3.16,5.01,7.94,12.59\right]  $, so that $\widehat{\delta}\left(
\widehat{t}\right)  =\widehat{\delta}_{0}-\widehat{r}\left(  \widehat
{t}-\widehat{t}_{0}\right)  $. Experimentally, this coincides with (i)
indenting the viscoelastic substrate up to a prescribed indentation depth,
(ii) waiting for a long dwell time so that the substrate fully relaxes, then
(iii) unloading with a given (constant) displacement rate. The normalized
unloading curves are shown in Fig. \ref{FigUnload} in terms of mean normal
stress versus remote indentation. As expected for a viscoelastic contact
problem, the unloading rate strongly influences the unloading trajectory. For
fast unloading, the substrate is stiff, hence a high load $\left(
K_{I}\varpropto P\right)  $ is required for the crack to advance. The  two
linear limit behaviour on unloading in Fig. \ref{FigUnload} correspond to
instantaneous and relaxed moduli, respectively. Clearly, pull-off stress
increases with rate of unloading, but the dependence is non monotonic for the
work of separation per unit area $\widehat{w}_{sep}=\frac{w_{sep}}%
{\alpha\sigma_{0}h_{0}}=-\int_{\widehat{\delta}_{P=0}}^{-\infty}%
\widehat{\sigma}d\widehat{\delta}$ which is proportional to the area
underneath the unloading curves ($\alpha={9\sqrt{3}}/{16}$, $\widehat{\delta
}_{P=0}$ is the indentation depth when the mean stress (or the load) vanishes).

\begin{figure}[!h]
\begin{center}
\includegraphics[width=4.38in]{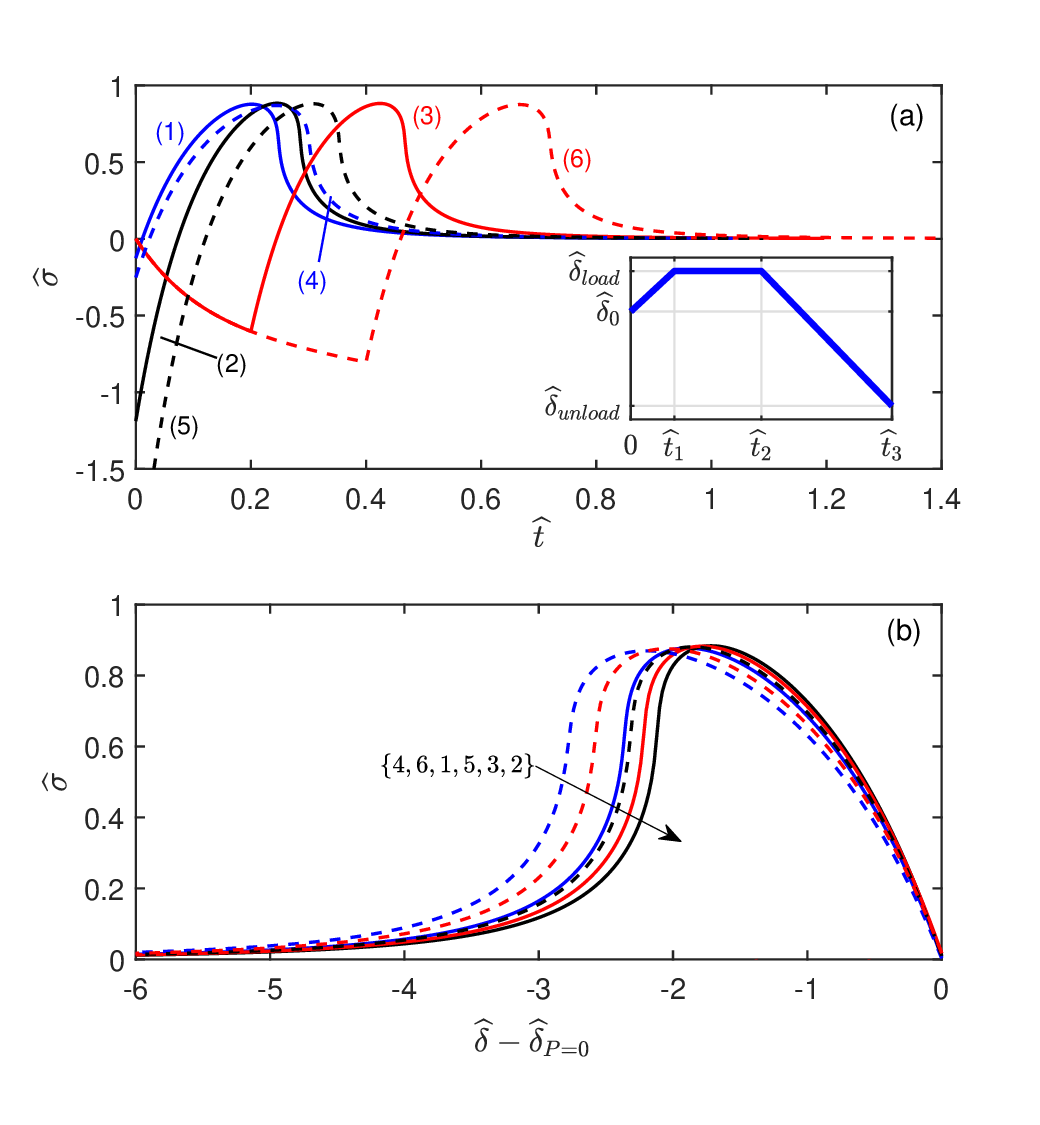}
\end{center}
\caption{(a) Unloading curves for $\Sigma_{0}=0.05$, $k=0.1$, punch of radius
$\widehat{a}/\widehat{a}_{0}=2.02$,\ unloaded from a fully relaxed
viscoelastic substrate starting from $\widehat{\delta}_{0}=1$ at different
unloading rates $\widehat{r}=\left[  0.13,3.16,5.01,7.94,12.59\right]  $. (b)
The same curves shown in panel (a) are reported here after shifting the
horizontal axis by $\widehat{\delta}_{P=0}$, which is the indentation depth at
which the normal stress vanishes during unloading.}%
\label{FigUnload3}%
\end{figure}

The response of a viscoelastic material is in general "history-dependent", as
from the hereditary integral in Eq. (\ref{int2}). Curves in Fig.
\ref{FigUnload} were obtained unloading the flat punch from a fully relaxed
substrate. To investigate the effect of the loading history, we run several
numerical simulations, with different loading protocols, but fixing the
unloading rate at $\widehat{r}=10$. The simulations were conducted in
displacement control and for $\widehat{\delta}\left(  \widehat{t}\right)  $ we
assumed a trapezoidal function, whose key parameters are defined in the inset
of Fig. \ref{FigUnload3}a. We introduce the dwell time $\widehat{t}%
_{dwell}=\widehat{t}_{2}-\widehat{t}_{1}$, the loading rate $\widehat
{r}_{load}=\left(  \widehat{\delta}_{load}-\widehat{\delta}_{0}\right)
/\widehat{t}_{1}$ and the unloading rate $\widehat{r}=\left(  \widehat{\delta
}_{load}-\widehat{\delta}_{unload}\right)  /\left(  \widehat{t}_{3}%
-\widehat{t}_{2}\right)  $. Results in Fig. \ref{FigUnload3}ab for each
unloading curve correspond to the parameters in Table 1. We considered the
cases when the punch is unloaded: (i) after very slow loading from a fully
relaxed substrate (blue curves, (1,4)), (ii) after very rapid loading so that
the substrate appears elastic with $E\left(  t=0\right)  =E_{\infty}$ (black
curves, (2,5)), (iii) after indenting the substrate at a constant loading rate
$\widehat{r}_{load}=5$ (red curves, (3,6)). Notice that for the curves
$\left(  1,2,4,5\right)  $ the loading phase is not shown. The maximum
indentation depth $\widehat{\delta}_{load}$ was fixed equal to $\widehat
{\delta}_{load}=1$ for the curves $\left(  1,2,3\right)  $ (solid lines) and
$\widehat{\delta}_{load}=2$ for the curves $[4,5,6]$ (dashed lines). Despite
the very different unloading trajectories, one notices that a key quantity
such as the pull-off stress (Fig. \ref{FigUnload3}ab)\ is almost unaffected by
the loading history. Fig. \ref{FigUnload3}b shows the same curves reported in
\ref{FigUnload3}a after shifting the horizontal axis by $\widehat{\delta
}_{P=0}$, which is the indentation depth at which the normal load vanishes
during unloading, so that one can better appreciate the slight changes in the
unloading trajectories. In the rest of the paper, unless differently stated,
we will assume $\widehat{\delta}_{0}=\widehat{\delta}_{load}=1$ and
$\widehat{t}_{dwell}=0.$

\begin{center}%
\begin{tabular}
[c]{|c|c|c|c|c|c|}\hline
Curve & $\widehat{\delta}_{0}$ & $\widehat{\delta}_{load}$ & $\widehat
{r}_{load}$ & $\widehat{r}$ & $\widehat{t}_{dwell}$\\\hline
\multicolumn{1}{|l|}{(1) - blue solid} & $1$ & $1$ & very slow & $10$ &
$0$\\\hline
\multicolumn{1}{|l|}{(2) - black solid} & $1$ & $1$ & very fast & $10$ &
$0$\\\hline
\multicolumn{1}{|l|}{(3) - red solid} & $0$ & $1$ & $5$ & $10$ & $0$\\\hline
\multicolumn{1}{|l|}{(4) - blue dashed} & $2$ & $2$ & very slow & $10$ &
$0$\\\hline
\multicolumn{1}{|l|}{(5) - black dashed} & $2$ & $2$ & very fast & $10$ &
$0$\\\hline
\multicolumn{1}{|l|}{(6) - red dashed} & $0$ & $2$ & $5$ & $10$ & $0$\\\hline
\end{tabular}

Table 1 - Set of parameters that defines the loading protocol of the curves
shown in Fig. \ref{FigUnload3}ab. Notice that for all the simulations the
unloading rate is fixed at $\widehat{r}=10$ and $\widehat{t}_{dwell}=0$ for
all the simulations.
\end{center}

\bigskip

\begin{figure}[h]
\begin{center}
\includegraphics[width=4.3799in]{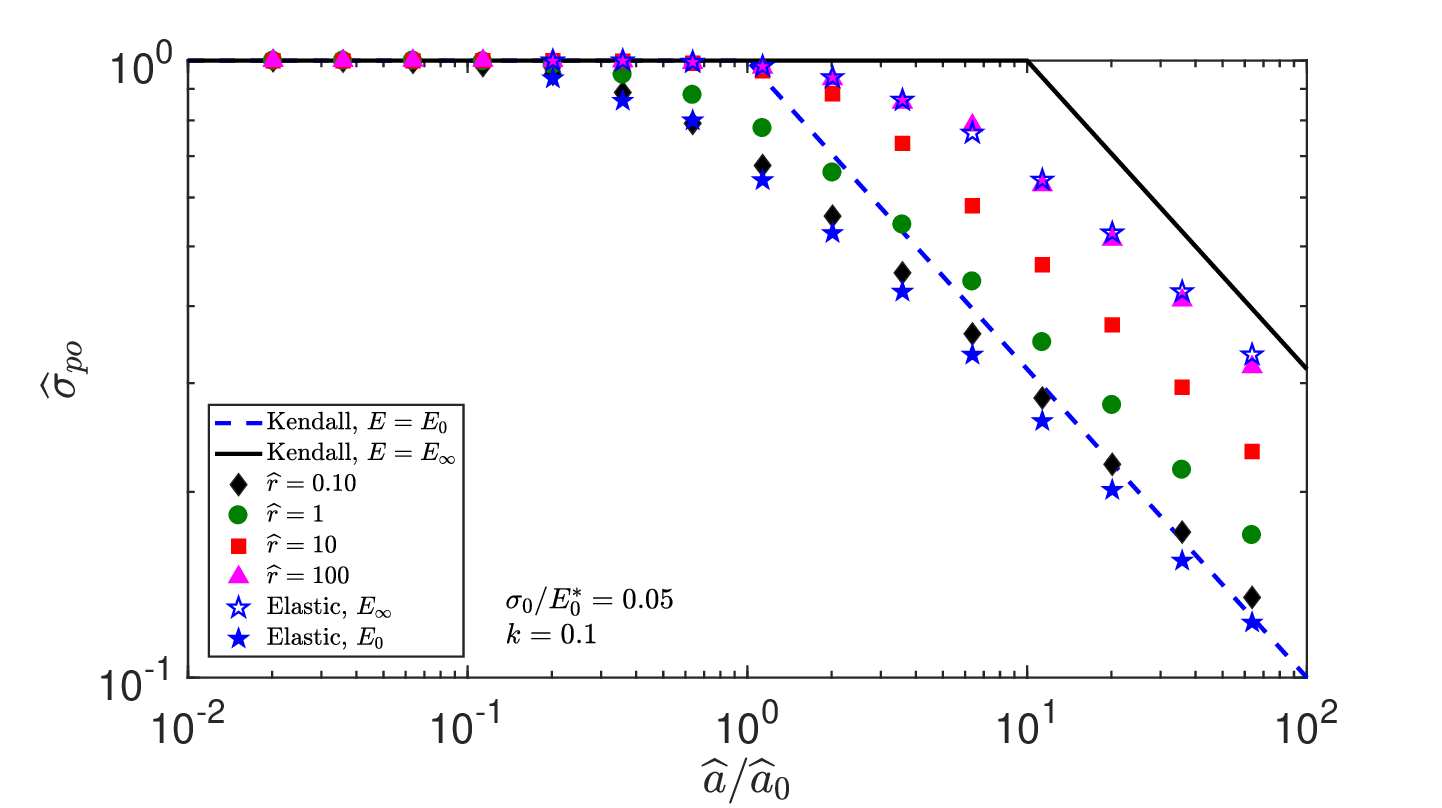}
\end{center}
\caption{Normalized pull-off stress as a function of the normalized contact radius for $\Sigma_{0}=0.05$, $k=0.1$. Unloading starts from $\widehat{\delta}_{0}=\widehat{\delta}_{load}=1$ from a fully relaxed substrate and is
performed at a constant unloading rate $\widehat{r}=\left[
0.1,1,10,100\right]  $, respectively black diamonds, green circles, red
squares and purple triangles. The LEFM limits for slow and fast unloading rate
are shown respectively as blue dashed and black solid lines. {Filled (empty) stars show the results obtained unloading an \textit{elastic} substrate with $E=E_0$ ($E=E_{\infty}$).}}%
\label{FigKendall}%
\end{figure}

Having assessed that the loading history does not influence the pull-off
stress $\widehat{\sigma}_{po}$, in Fig. \ref{FigKendall} we investigated the
variation of $\widehat{\sigma}_{po}$ as a function of the normalized punch
radius for four unloading rates $\widehat{r}=\left[  0.1,1,10,100\right]  $
(respectively black diamonds, green circles, red squares and pink triangles)
and starting from a fully-relaxed substrate. For $\widehat{a}/\widehat{a}%
_{0}\ll1$ the cohesive limit is reached, where the pull-off stress is
independent on both the loading rate and the punch radius and it approaches
the theoretical value $\widehat{\sigma}_{po}=1$. For $\widehat{a}/\widehat
{a}_{0}\gg1$ the curves follow the square root scaling imposed by LEFM, the
pull-off stress increases with the unloading rate and the pull-off data remain
bounded by the "slow" and "fast" limits dictated by the Kendall
\cite{Kendall1971} solution (Eq. (\ref{slowlimit},\ref{fastlimit})),
respectively blue-dashed and solid black lines.
{Notice that numerically we had to accomodate a
small cohesive zone before we find the maximum force which sligthly reduced the
peak force from the theoretical Kendall limit (see Fig. \ref{FigKendall}).}

\begin{figure}[t!]
\begin{center}
\includegraphics[width=4.3in]{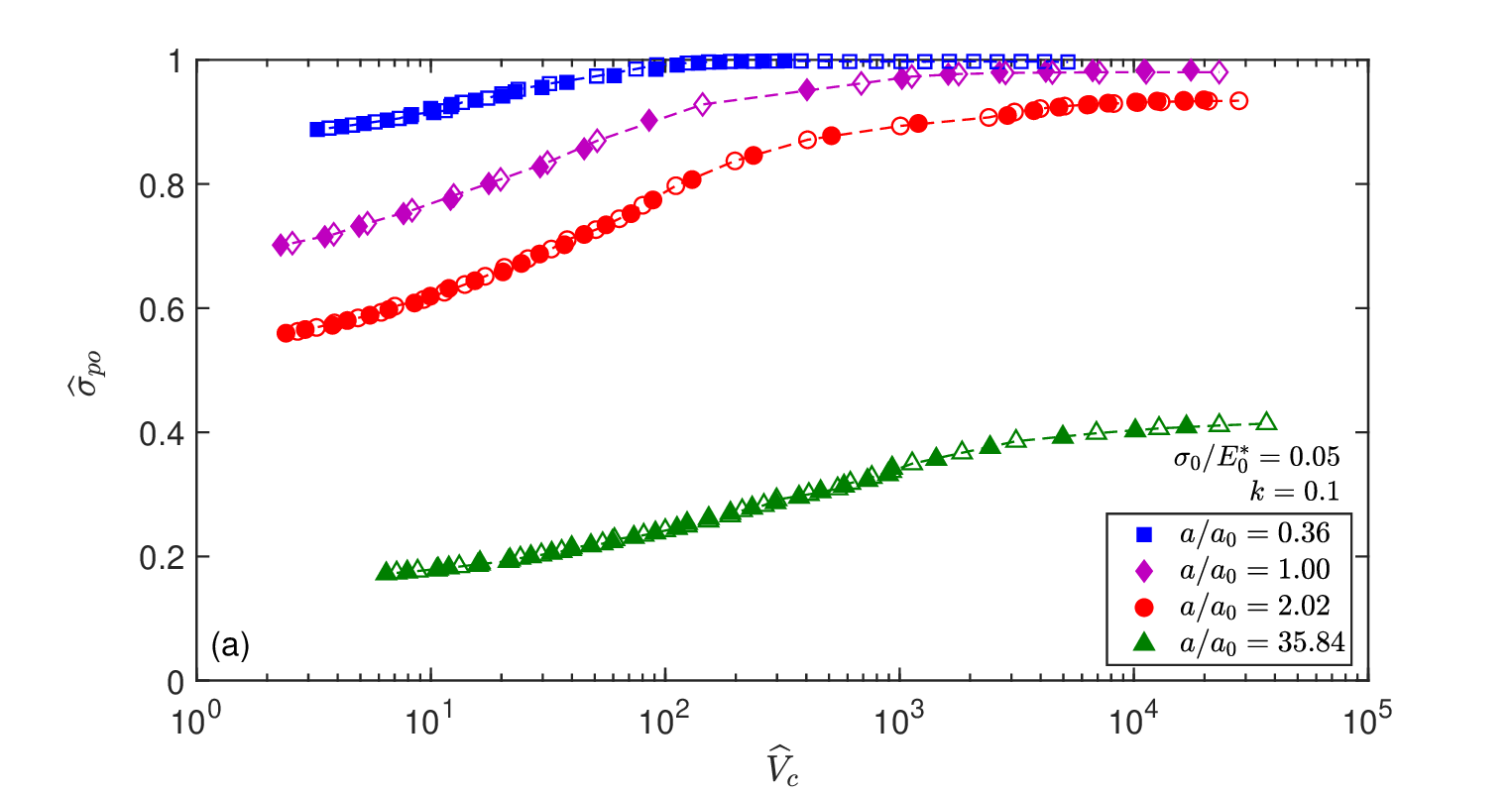}
\includegraphics[width=4.3in]{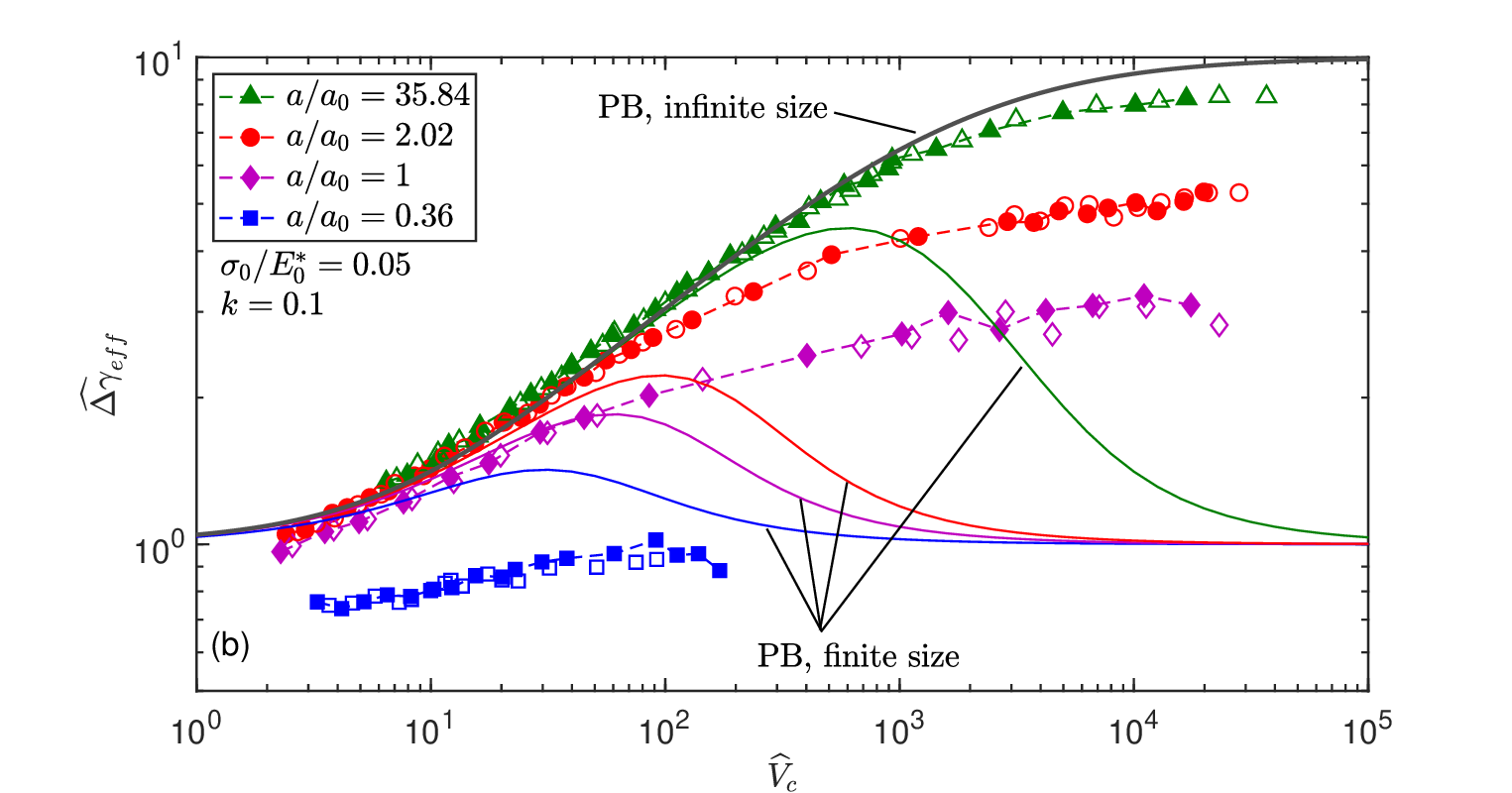}
\end{center}
\caption{{(a) Normalized pull-off stress as a function of the normalized
unloading rate. (b) Normalized effective work of adhesion as a function of the crack velocity at pull-off $\widehat{V}_{c}$. In panel (b) solid lines refer to Persson and Brener theory
\cite{Persson2005} as extended for finite size systems by
Persson \cite{persson2017} slightly simplified by Ciavarella and Papangelo 
(Eq. (17) in \cite{CiaPap2021}), but with a shift on the crack
velocity. The same colour corresponds to the same ratio $\widehat{a}%
/\widehat{a}_{0}$. The grey solid line refers to Persson and Brener theory for
infinite systems (see Eq. (14) in Ref. \cite{CiaPap2021}). In both panels
$\Sigma_{0}=0.05$, $k=0.1$, $\widehat{\delta}_{0}=\widehat{\delta}_{load}=1$,
$\widehat{a}/\widehat{a}_{0}=\left[0.36,1,2.02,35.84\right]  $ respectively
blue squares, purple diamonds, red circles, green triangles. Filled symbols are used for data
obtained unloading the punch form a fully relaxed substrate, while empty symbols refer to data obtained
after very fast loading.}}%
\label{Figgeff}%
\end{figure}

The variation of the pull-off stress as a function of the crack speed at
pull-off $\widehat{V}_{c}=V_{c}\tau/h_{0}$ is shown in Fig. \ref{Figgeff}a,
for four different punch radii $\widehat{a}/\widehat{a}_{0}=\left[
0.36,1.00,2.02,35.84\right]  $ (respectively blue, purple, red and green curves) unloading
either after a very slow loading phase (filled symbols, when unloading starts $E\simeq E_0$) or after a very fast
loading phase (empty symbols, when unloading starts $E\simeq E_\infty$) and for $\Sigma_{0}=0.05$, $k=0.1$,
$\widehat{\delta}_{load}=1$. As suggested in Fig. \ref{FigUnload3}, the
pull-off stress is not affected by the loading history, moreover for small
radii (blue curve) a change of $3$ orders of magnitude of the unloading rate
resulted in a variation of the pull-off stress by only a $10\%,$ as we are
near the cohesive strength limit. Notice that we did not find an appreciable
dependence of the pull-off stress on $\widehat{\delta}_{load}$ (not shown).

{The same data shown in Fig. \ref{Figgeff}a are also reported in Fig.
\ref{Figgeff}b, this time plotted as a normalized effective work of adhesion
$\widehat{\Delta\gamma}_{eff}=\Delta\gamma_{eff}/\Delta\gamma$ versus the
crack speed at pull-off $\widehat{V}_{c}$, being $\Delta\gamma_{eff}%
=P_{po}^{2}/8\pi E_{0}^{\ast}a_{c}^{3}$, or, in dimensionless form $\widehat
{\Delta\gamma}_{eff}=\frac{2\pi}{9\sqrt{3}}\widehat{\sigma}_{po}^{2}\widehat{a} (\frac{\widehat{a}}{\widehat{a}_c})^3
\Sigma_{0}$. The numerical results are compared with the prediction of Persson
and Brener theory \cite{Persson2005} as extended for finite size systems (where finite size is implied by the difference between the size of crack ligament $a_c$ and the finite punch radius $a$) by
Persson \cite{persson2017} slightly simplified by Ciavarella and Papangelo
(Eq. (17) in Ref. \cite{CiaPap2021}). Notice that in order to have a
reasonable agreement with the numerical results, we had to use a
shift factor on the crack velocity definition\footnote{In Eq. (17) of Ref. \cite{CiaPap2021} we used as
a reference velocity $v_{0,new}=2\pi^2 v_0$, being $v_{0}=\frac{\Delta\gamma
E_{0}^{*}}{(2\pi \sigma_{0})^{2}\tau}$.} \cite{CiaPap2021,CCM2021}.} {The same color corresponds
to the same ratio $\widehat{a}/\widehat{a}_{0}$, while the grey curve is the
reference to an infinite size system. The numerical results show that the
effective work of adhesion increases monotonically with the crack speed in all
cases. In particular, for large systems $\widehat{a}/\widehat{a}_{0}=35.84$
LEFM holds, hence for slow crack velocity $\widehat{\Delta\gamma}_{eff}%
\simeq1$, while at high velocity, we obtained an enhancement close to $E_\infty/E_0$.} For smaller radii, finite size effects come at play, but $\widehat{\Delta\gamma}_{eff}$ still increases monotonically with the crack
speed. Moreover, for small radii, the normalized effective work of adhesion
can be smaller than unity. This should not surprise as, even for a purely
elastic problem, the pull-off stress is less than that expected from LEFM for
small radii due to the transition towards the cohesive limit region which is
equivalent to a reduced work of adhesion. In general, while we can say that
there is a quantitative agreement with dissipation based theories for large
radii, the results contrast both quantitatively and qualitatively for finite
size, and in particular, there is no evidence of a maximum of load at
intermediate speeds. 

\begin{figure}[t]
\begin{center}
\includegraphics[width=4.3791in]{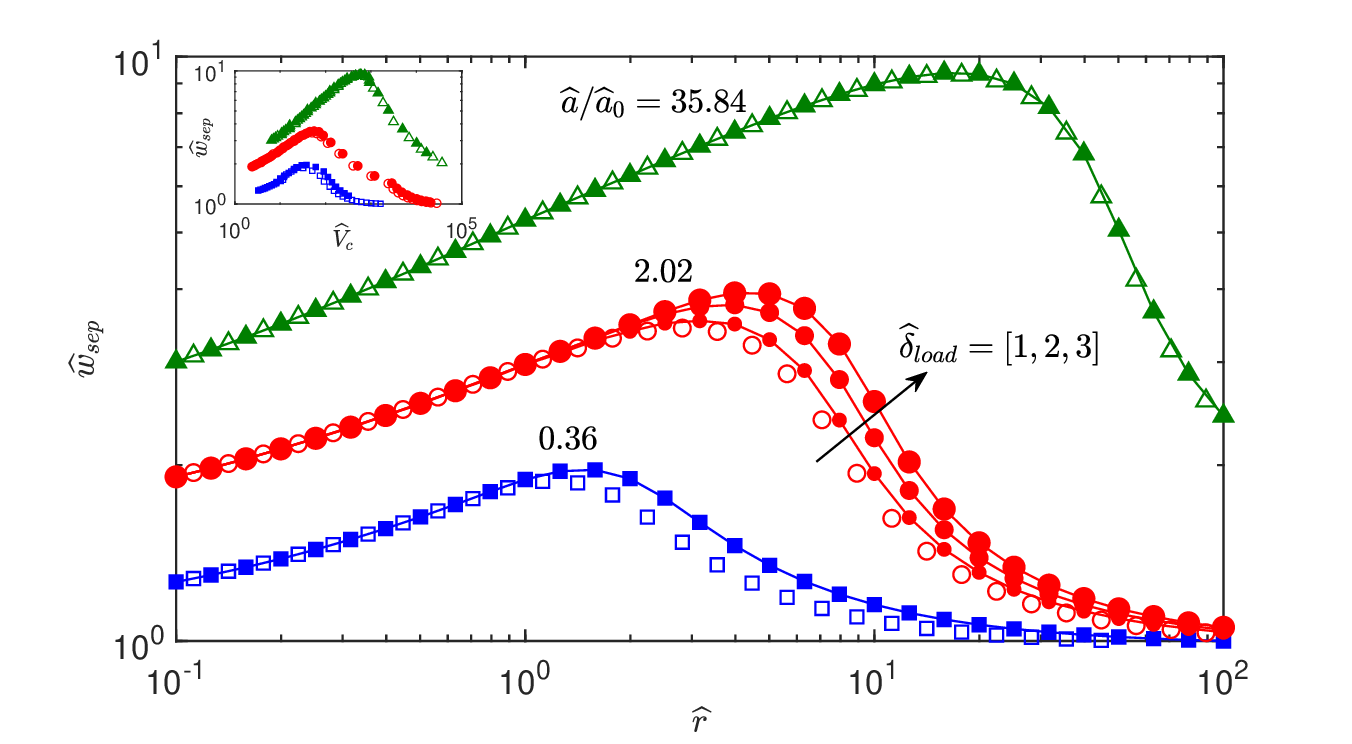}
\end{center}
\caption{Normalized work of separation as a function of the normalized
unloading rate for ${\Sigma}_{0}=0.05$, $k=0.1$, $\widehat{\delta}%
_{0}=\widehat{\delta}_{load}=1$, $\widehat{a}/\widehat{a}_{0}=\left[
0.36,2.02,35.84\right]  $ respectively blue squares, red circles, green
triangles. Filled symbols refer to the case when unloading starts after very
slow loading, while empty symbols refer to pull-off data obtained after very
fast loading. For the case $\widehat{a}/\widehat{a}_{0}=2.02$ and unloading
from fully relaxed substrate we tested several values of $\widehat{\delta}%
_{0}=\widehat{\delta}_{load}=\left[  1,2,3\right]  $ (the circle dimension
increases with $\widehat{\delta}_{load}$). In the inset the work of separation
is shown as a function of the crack velocity (only the data referring to
$\widehat{\delta}_{0}=\widehat{\delta}_{load}=1 $ are shown).}%
\label{Figwsep}%
\end{figure}

A key quantity to investigate is the work of separation $\widehat{w}_{sep}$,
as it gives an estimate of the energy that is dissipated during the unloading
phase. We obtained this quantity as a function of the normalized unloading
rate for $\widehat{\delta}_{0}=\widehat{\delta}_{load}=1$, $\widehat
{a}/\widehat{a}_{0}=\left[  0.36,2.02,35.84\right]  $ respectively blue
squares, red circles and green triangles (Fig. \ref{Figwsep}). Filled symbols
refer to the case when the punch is unloaded after a very slow loading, while
empty symbols refer to data obtained unloading after a very fast loading
phase. The inset shows the work of separation as a function of the crack speed.

Notice that, contrary to the case of the pull-off stress and as suggested by
Fig. \ref{FigUnload3}, the work of separation shows a weak dependence on the
loading rate, limited to small punch radius $\left(  \widehat{a}%
/\widehat{a}_{0}=0.36\right)  $ and intermediate unloading rates. For the case
$\widehat{a}/\widehat{a}_{0}=2.02$, and unloading from a fully relaxed
substrate, several values of $\widehat{\delta}_{0}=\widehat{\delta}%
_{load}=\left[  1,2,3\right]  $ were tested, whose results are shown as red
circles with the circle dimension increasing with $\widehat{\delta}_{load}$.
For intermediate rates, we find a slight dependence of $\widehat{w}_{sep}$ on
$\widehat{\delta}_{load}$, which makes the picture a bit more elaborate than
what happens for the pull-off stress. Importantly, while the pull-off stress
increases monotonically with respect to the unloading rate, independently on
the loading history, the work of separation retains a bell shape. Notice that
for small punch radii $\left(  \widehat{a}/\widehat{a}_{0}<<1\right)  $ the
elastic limit at high and low velocity is $\widehat{w}_{sep}=1$, as we have a
progressive detachment of the interface with uniform stress up to very large
displacements, while in the LEFM region $\left(  \widehat{a}/\widehat{a}%
_{0}>>1\right)  $ the elastic limit at high and low velocity is $\widehat
{w}_{sep}=2$, because there is essentially a linear load-displacement curve up
to the "brittle" rupture at the Kendall load and it can be easily shown that
the gives $\widehat{w}_{sep}=2$, so that some strain energy must be released
at the point of fracture. Hence, the problems are in all respects elastic both
at very low and very high speeds, and dissipation is not the cause of the load
enhancement at fast unloading, contrary to the classical de Gennes picture. The
load increases of the ratio instantaneous to relaxed moduli, but the remote
displacement decreases by the same ratio, hence the resulting area integral is
the same at very low unloading rates, and at very high ones.

\section{Extension of Maugis cohesive zone model}

In this section, we attempt a generalization of the elastic cohesive model of
Maugis \cite{Maugis} (see also the work of Tang \& Hui \cite{TangHui}) to the
case of viscoelasticity. The problem of unloading a rigid axisymmetric flat
punch from an elastic half-space of composite elastic modulus $E^{\ast}$ has
been studied using a Dugdale cohesive law \cite{Maugis,TangHui} which showed
that the peeling force $P$ is related to the radius of the ligament of the crack
$a_{c}$ as follows%

\begin{equation}
P=2a_{c}\sigma_{0}\left[  \sqrt{a^{2}-a_{c}^{2}}+\frac{a^{2}}{a_{c}}\cos
^{-1}\frac{a_{c}}{a}\right]  \label{PpoMD}%
\end{equation}
where we recall $a$ is the punch radius, while {$a_c$ is the radius of the ligament of the crack which is a function of time}, hence the cohesive stresses may exist
in the annulus $a_{c}\leq r\leq a$. The pull-off occurs when the separation at
the crack mouth $r=a$ reaches the Critical Opening Distance (COD) $h_{c}$ in
the Dugdale force-separation law, as any smaller radius would lead to a
smaller force. Considering the crack profile this condition translates into
\cite{TangHui}%

\begin{equation}
h_{c}=\frac{4\sigma_{0}}{\pi E^{\ast}}\left[  a_{c}-a+\sqrt{a^{2}-a_{c}^{2}%
}\cos^{-1}\frac{a_{c}}{a}\right]  \label{acpo}%
\end{equation}
where equating the surface energy $\Delta\gamma$ in a LJ force-separation law
with that of a Dugdale model and having the same theoretical strength
$\sigma_{0}$, gives $h_{c}=\alpha h_{0}\simeq0.974h_{0}$. Solving Eq.
(\ref{acpo}) for $a_{c}$ one determines the critical radius of the ligament of the crack at pull-off
and substituting into Eq. (\ref{PpoMD}) the pull-off force is obtained. In the
dimensionless formulation introduced before the equation for the normal stress
is unaffected by the modulus%

\begin{equation}
\widehat{\sigma}=\frac{2}{\pi}\frac{\widehat{a}_{c}}{\widehat{a}}\left[
\sqrt{1-\left(  \frac{\widehat{a}_{c}}{\widehat{a}}\right)  ^{2}}%
+\frac{\widehat{a}}{\widehat{a}_{c}}\cos^{-1}\frac{\widehat{a}_{c}}%
{\widehat{a}}\right]  \label{PullCOD}%
\end{equation}
while the condition for the COD is modified for a viscoelastic half-space, by
substituting the inverse of the elastic modulus with an effective compliance
$c_{eff}\left(  \widehat{t}_{b}\right)  =E_{0}^{\ast}C\left(  t_{b}\right)  $,
with $t_{b}$ a characteristic time%

\begin{equation}
\frac{4\Sigma_{0}}{\pi}c_{eff}\left(  \widehat{t}_{b}\right)  \left[
\widehat{a}_{c}-\widehat{a}+\widehat{a}\sqrt{1-\left(  \frac{\widehat{a}_{c}%
}{\widehat{a}}\right)  ^{2}}\cos^{-1}\frac{\widehat{a}_{c}}{\widehat{a}%
}\right]  -\frac{9\sqrt{3}}{16}=0 \label{COD}%
\end{equation}

\begin{figure}[t]
\begin{center}
\includegraphics[width=4.3799in]{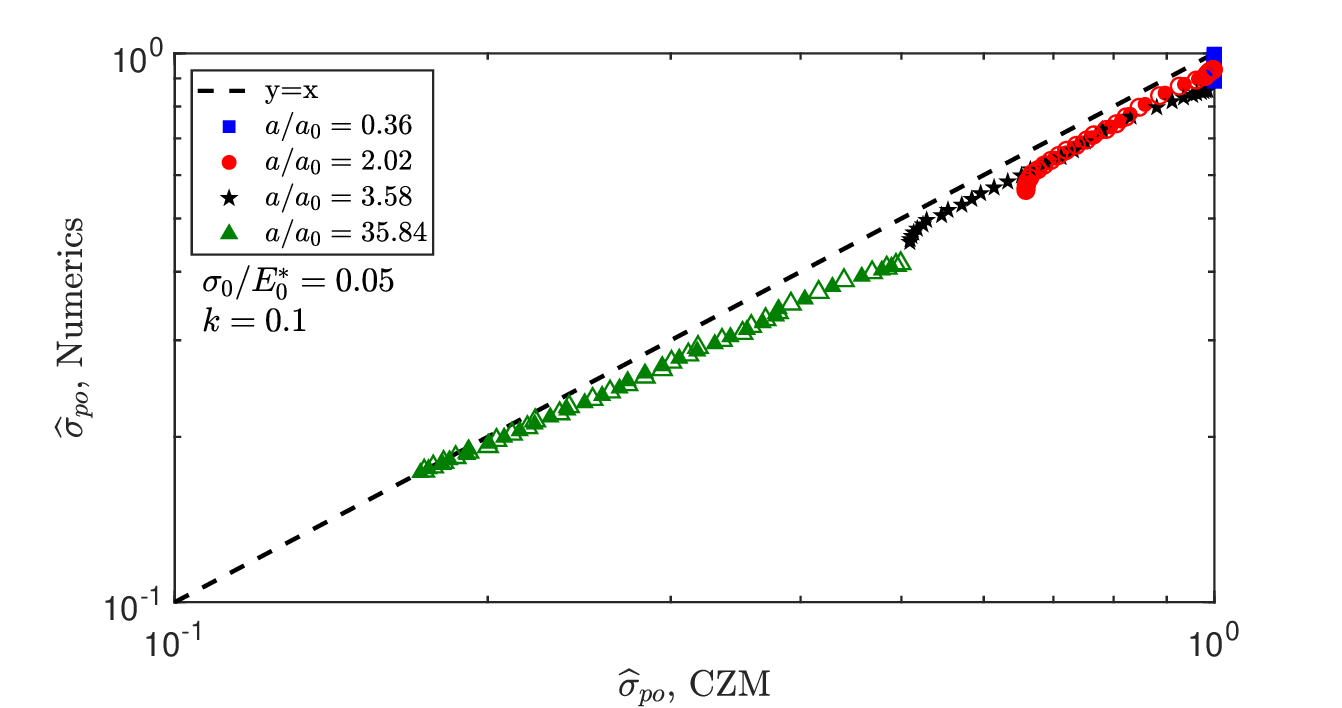}
\end{center}
\caption{The pull-off stress obtained with BEM\ numerical simulations is
compared with that predicted by the CZM (Eq. (\ref{Ceff},\ref{COD}%
,\ref{PullCOD})) for the following set of parameters $\Sigma_{0}=0.05$,
$k=0.1$, $\widehat{\delta}_{0}=\widehat{\delta}_{load}=1$ and $\widehat
{a}/\widehat{a}_{0}=\left[  0.36,2.02,3.58,35.84\right]  $, respectively blue
squares, red circles, black stars, green triangles. Filled (Empty) symbols
refer to the case when unloading starts from a relaxed (unrelaxed) substrate.}%
\label{Figpull}%
\end{figure}

In analogy with the suggestion of Schapery, we define $t_{b}\simeq b/V$ being
$b$ the width of the cohesive zone and $V$ the crack
velocity\footnote{Schapery uses $t_{b}=b/3V$, for a constant stress (Dugdale)
model.}. For semi-infinite crack, and a Dugdale cohesive stress law, Schapery found%

\begin{equation}
b=\frac{\pi}{8}\left(  \frac{K_{I}}{\sigma_{0}}\right)  ^{2}=\frac{\pi^{2}%
}{32}\left(  \frac{\sigma}{\sigma_{0}}\right)  ^{2}\left(  \frac{a}{a_{c}%
}\right)  ^{4}a_{c} \label{bS}%
\end{equation}
where we assumed, during propagation, the mode-I SIF is given by
$K_{I}=P/\sqrt{4\pi a_{c}^{3}}$. In the numerical simulation, we used a
standard linear viscoelastic material, hence%

\begin{equation}
c_{eff}\left(  \widehat{t}_{b}\right)  =1+\left(  k-1\right)  \exp\left[
-\frac{\pi^{2}}{96}\frac{\widehat{\sigma}^{2}\widehat{a}_{c}}{\widehat{V}%
}\left(  \frac{\widehat{a}}{\widehat{a}_{c}}\right)  ^{4}\right]  \label{Ceff}%
\end{equation}

\begin{figure}[t]
\begin{center}
\includegraphics[width=4.3799in]{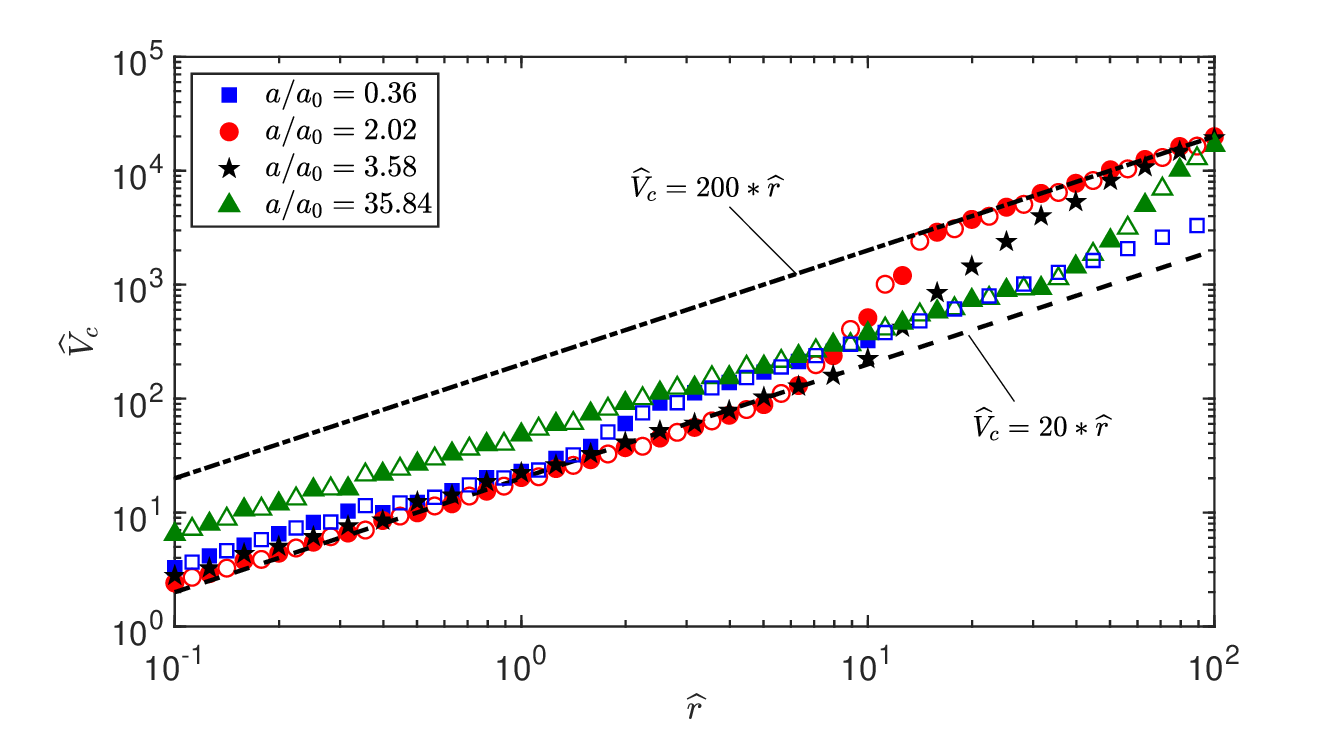}
\end{center}
\caption{Crack propagation velocity at pull-off $\widehat{V}_{c}$ as a
function of the unloading rate $\widehat{r}$ for the same data as in Fig.
(\ref{Figpull}, the same symbols are used). }%
\label{FigVc}%
\end{figure}

At every time-step we determined the crack position based on the location of
the maximum adhesive stress at the interface, hence we estimated $V_{c}$, the
crack velocity at pull-off, and used this value in the Cohesive Zone Model
(CZM, Eq. (\ref{Ceff},\ref{COD},\ref{PullCOD})) to determine the theoretical
pull-off stress which was then compared with that obtained numerically in Fig.
\ref{Figpull} for a set of punch radii $\widehat{a}/\widehat{a}_{0}=\left[
0.36,2.02,3.58,35.84\right]  $ (respectively blue squares, red circles, black
stars, green triangles) {unloading from a fully relaxed substrate (filled
symbols, $E\simeq E_0$ when unloading starts) or from a not-relaxed substrate (empty symbols, $E\simeq E_\infty$ when unloading starts).} Although small
discrepancies appear with respect to the numerical data, the CZM provides a
fairly good estimate of the pull-off stress, with deviations not larger than a
$10\%$ between numerical and theoretical data.

As with semi-infinite crack theories, we have described the model as a
function of the crack speed, which, in an adhesive experiment, is unlikely to
be constant and has to be measured in numerical/real experiments, where
generally the remote displacement rate is imposed. The relation between the
external unloading rate $\widehat{r}$ with the crack speed at pull-off
$\widehat{V}_{c}$ is shown in Fig. \ref{FigVc} for the same data as in Fig.
\ref{Figpull}. For small punch radii (blue squares) the behaviour is almost
linear and a small deviation can be seen only around $\widehat{r}\approx2$,
while for $\widehat{a}/\widehat{a}_{0}>1$ we clearly found two linear regimes
and a transition zone between the two linear dashed lines shown in Fig.
\ref{FigVc}, which perhaps may be used to roughly estimate the crack speed as
a function of the loading rate.

\section{Conclusions}

In this work, we have studied the adhesive behaviour of a circular rigid flat
punch that is placed in contact with a viscoelastic half-space. We showed that
depending on the punch radius there exists a transition from a cohesive limit,
where debonding happens at a uniform stress equal to the theoretical strength
of the interface, to a regime that is governed by classical LEFM concepts,
extended to the viscoelastic case. Our study has focused on two key
quantities, such as the pull-off stress and the work of separation. We found
that the pull-off stress is a monotonically increasing quantity of the
unloading rate and is almost independent on the contact history. This is
markedly different from the case of a Hertzian indenter which has been
extensively studied \cite{AffVio2022,VioAff2022size,VioAffRange} and showed a
strong influence on the preload. Indeed, we suggest here that the dependence on the loading
history of the pull-off stress is geometry dependent. In our numerical
simulations we also measured the work of separation $w_{sep}$ as the energy
per unit area that is dissipated to separate the contact. Indeed, in an
attempt to interpret the maximum of \textit{load} (and resulting instability)
in the peeling experiments of Gent and Petrich \cite{gentpetrich}, de Gennes
\cite{deGennes} proposed a theory of viscoelastic semi-infinite crack
propagation based on \textit{dissipation} away from the crack tip reaching a
maximum at intermediate speeds. We found that $w_{sep}$ has indeed a maximum
at intermediate crack speed, but this does not correspond to a maximum of the
load and hence dissipated energy cannot be directly used to determine the
debonding load. {Notice that in our geometry, where the flat punch is rigid, the
stress intensity factor $K_{I}$ doesn't need any geometrical
correction factor as for example introduced in \cite{Vio2023} as it
is always equal to $K_{I}=P/\sqrt{4\pi a_{c}^{3}}$ where $a_{c}$
is initially the punch radius but then becomes the current size of
the ligament of the crack. Hence our numerical results cannot be explained by a ``corrected'' Persson-Brener theory for infinite systems as suggested by Violano et al. \cite{Vio2023} for a finite viscoelastic plate geometry.}
Finally, we have proposed a cohesive zone model for debonding
of a flat punch from a viscoelastic substrate, stemming from the elastic
theory by Maugis \cite{Maugis} (see also \cite{TangHui}), which proved to be
reasonably good in terms of pull-off stress prediction. Unfortunately, in
general we cannot impose externally the crack speed, but the remote
displacement rate, and hence the theory only interprets data but is not predictive.

\section*{Statements \& Declarations}

\subsection*{Funding}

A.P. was supported by the European Union (ERC-2021-STG, “Towards Future Interfaces With Tuneable Adhesion By Dynamic Excitation” — SURFACE, Project ID: 101039198, CUP: D95F22000430006). Views and opinions expressed are however those of the authors only and do not necessarily reflect those of the European Union or the European Research Council. Neither the European Union nor the granting authority can be held responsible for them. A.P. was partially supported by Regione Puglia (Italy), project ENOVIM (CUP: D95F21000910002) granted within the call ”Progetti di ricerca scientifica innovativi di elevato standard internazionale” (art. 22 della legge regionale 30 novembre 2019, n. 52 approvata con A.D. n. 89 of 10-02-2021, BURP n. 25 del 18-02-2021). A.P. and M.C. acknowledge support by the Italian Ministry of University and Research (MUR) under the programme “Department of Excellence” Legge 232/2016 (Grant No. CUP - D93C23000100001). 

\subsection*{Data availability}

The dataset generated for this article is availbale at DOI: 10.5281/zenodo.7757174 

\subsection*{Competing Interests}

The authors have no relevant financial or non-financial interests to disclose.

\subsection*{Author Contributions}

A.P. and M.C. contributed equally to this work.

\subsection*{Published Version}

{This article was published in Tribology Letters, Springer, DOI: 10.1007/s11249-023-01720-9}


\begin{thebibliography}{99}                                                                                               %


\bibitem {Gio2021a}Giordano, G., Carlotti, M., \& Mazzolai, B. (2021). A
Perspective on Cephalopods Mimicry and Bioinspired Technologies toward
Proprioceptive Autonomous Soft Robots. Advanced Materials Technologies, 6(12), 2100437.

\bibitem {Mazzolai2019}Mazzolai, B., Mondini, A., Tramacere, F., Riccomi, G.,
Sadeghi, A., Giordano, G., ... \& Carminati, S. (2019). Octopus-Inspired Soft
Arm with Suction Cups for Enhanced Grasping Tasks in Confined Environments.
Advanced Intelligent Systems, 1(6), 1900041.

\bibitem {Gio2021b}Giordano, G., Gagliardi, M., Huan, Y., Carlotti, M.,
Mariani, A., Menciassi, A., ... \& Mazzolai, B. (2021). Toward Mechanochromic
Soft Material-Based Visual Feedback for Electronics-Free Surgical Effectors.
Advanced Science, 8(15), 2100418.

\bibitem {VCacu}Cacucciolo, V., Shintake, J., Kuwajima, Y., Maeda, S.,
Floreano, D., \& Shea, H. (2019). Stretchable pumps for soft machines. Nature,
572(7770), 516-519.

\bibitem {Shui}Shui, L., Jia, L., Li, H., Guo, J., Guo, Z., Liu, Y., ... \&
Chen, X. (2020). Rapid and continuous regulating adhesion strength by
mechanical micro-vibration. Nature communications, 11(1), 1-7.

\bibitem {Arzt}Arzt, E., Quan, H., McMeeking, R. M., \& Hensel, R. (2021).
Functional surface microstructures inspired by nature--from adhesion and
wetting principles to sustainable new devices. Progress in Materials Science, 100778.

\bibitem {Lunni2020}Lunni, D., Giordano, G., Pignatelli, F., Filippeschi, C.,
Linari, S., Sinibaldi, E., \& Mazzolai, B. (2020). Light-assisted
electrospinning monitoring for soft polymeric nanofibers. Scientific reports,
10(1), 1-12.

\bibitem {Lunni2018}Lunni, D., Giordano, G., Sinibaldi, E., Cianchetti, M., \&
Mazzolai, B. (2018, April). Shape estimation based on kalman filtering:
Towards fully soft proprioception. In 2018 IEEE International Conference on
Soft Robotics (RoboSoft) (pp. 541-546). IEEE.

\bibitem {Asbeck}Asbeck, A., Dastoor, S., Parness, A., Fullerton, L., Esparza,
N., Soto, D., ... \& Cutkosky, M. (2009). Climbing rough vertical surfaces
with hierarchical directional adhesion. In 2009 IEEE International Conference
on Robotics and Automation (pp. 2675-2680). IEEE.

\bibitem {Shintake}Shintake, J., Cacucciolo, V., Floreano, D., \& Shea, H.
(2018). Soft robotic grippers. Advanced Materials, 30(29), 1707035.

\bibitem {Dahlquist}Dahlquist, C. A. (1969). Pressure-sensitive adhesives. In
Treatise on Adhesion and Adhesives. Vol.2 (ed. R. L. Patrick), pp.219-260. New
York: Dekker.

\bibitem {Creton}Creton, C., \& Ciccotti, M. (2016). Fracture and adhesion of
soft materials: a review. \textit{Reports on Progress in Physics}, 79(4), 046601.

\bibitem {CPM2021}Ciavarella, M., Papangelo, A., \& McMeeking, R. (2021).
Crack propagation at the interface between viscoelastic and elastic materials.
Engineering Fracture Mechanics, 257, 108009.

\bibitem {Barquins1981}Barquins, M., \& Maugis, D. (1981). Tackiness of
elastomers. \textit{The Journal of Adhesion}, 13(1), 53-65.).

\bibitem {GentSchultz}Gent, A. N. and Schultz, J., (1972), Effect of wetting
liquids on the strength of adhesion of viscoelastic material, \textit{J.
Adhes.} 3(4), 281-294.

\bibitem {Gent}Gent, A. N., \& Petrich, R. P. (1969). Adhesion of viscoelastic
materials to rigid substrates. \textit{Proceedings of the Royal Society of
London. A. Mathematical and Physical Sciences}, 310(1502), 433-448.

\bibitem {Andrews}Andrews, E. H., \& Kinloch, A. J. (1974). Mechanics of
elastomeric adhesion. In \textit{Journal of Polymer Science}: Polymer Symposia
(Vol. 46, No. 1, pp. 1-14). New York: Wiley Subscription Services, Inc., A
Wiley Company.

\bibitem {Greenwood1981}Greenwood J A and Johnson K L (1981) The mechanics of
adhesion of viscoelastic solids \textit{Phil. Mag. A} 43 697--711

\bibitem {Maugis}Maugis, D. (1992). Adhesion of spheres: the JKR-DMT
transition using a Dugdale model. Journal of colloid and interface science,
150(1), 243-269..

\bibitem {Rivlin}R. Rivlin , Paint Technol. (1944), 9 , 215. DOI: 10.1007/978-1-4612-2416-7\_179

\bibitem {Williams}Williams, M. L.; Landel, R. F.; Ferry, J. D. (1955) The
Temperature Dependence of Relaxation Mechanisms in Amorphous Polymers and
Other Glass-Forming Liquids. \textit{Journal of the American Chemical
Society,} 77, 3701-3707.

\bibitem {gentpetrich}Gent, A. N., \& Petrich, R. P. (1969). Adhesion of
viscoelastic materials to rigid substrates. Proceedings of the Royal Society
of London. A. Mathematical and Physical Sciences, 310(1502), 433-448.

{
\bibitem {Ceglie}Ceglie, M., Menga, N., \& Carbone, G. (2022). The role of interfacial
friction on the peeling of thin viscoelastic tapes. Journal of the Mechanics
and Physics of Solids, 159, 104706.}

\bibitem {Rice}Rice, J. R. (1978). Mechanics of quasi-static crack growth (No.
COO-3084-63; CONF-780608-3). Brown Univ., Providence, RI (USA). Div. of Engineering.

\bibitem {Graham}Graham, G. A. C. (1969) Two extending crack problems in
linear viscoelasticity theory, \textit{Q. Appl. Math.}, 27, 497--507

\bibitem {Schapery}Schapery, R. A. (1975) A theory of crack initiation and
growth in viscoelastic media. \textit{Int. J. Fracture}, 11, (Part I) 141--59

\bibitem {SchaperyII}Schapery, R. A. (1975). A theory of crack initiation and
growth in viscoelastic media II. Approximate methods of analysis. \textit{Int.
J. Fracture}, 11(3), 369-388.

\bibitem {Knauss}Knauss WG (1973b) On the steady propagation of a crack in a
viscoelastic sheet: experiments and analysis. In: Kausch HH, Hassel IA, Jaffe
RE (eds) Deformation and fracture of high polymers. Proceedings of the
same-name 1972 conference in Kronberg, Germany. Plenum Press, New York,
London, pp 501--541

\bibitem {deGennes}de Gennes, P. G. (1996). Soft adhesives.\textit{ Langmuir},
12(19), 4497-4500.

\bibitem {Persson2005}Persson, B. N. J., \& Brener, E. A. (2005). Crack
propagation in viscoelastic solids. Physical Review E, 71(3), 036123.

\bibitem {hui}Hui, C. Y., Zhu, B., \& Long, R. (2022). Steady state crack
growth in viscoelastic solids: A comparative study. Journal of the Mechanics
and Physics of Solids, 159, 104748.

\bibitem {CiaPap2021}Ciavarella, M., \& Papangelo, A. (2021). Effects of
finite thickness on crack propagation in viscoelastic materials. Engineering
Fracture Mechanics, 248, 107703.

\bibitem {CiaPapMec2022}Ciavarella, M., Papangelo, A., \& McMeeking, R. (2022)
On transient and steady state viscoelastic crack propagation in a double
cantilever beam specimen. International Journal of Mechanical Sciences, 229,
107510. DOI: 10.1016/j.ijmecsci.2022.107510

\bibitem {CCM2021}Ciavarella, M., Cricr\`{\i}, G., \& McMeeking, R. (2021). A
comparison of crack propagation theories in viscoelastic materials.
Theoretical and Applied Fracture Mechanics, 116, 103113.

\bibitem {xuHui}Xu, D. B., Hui, C. Y., \& Kramer, E. J. (1992). Interface
fracture and viscoelastic deformation in finite size specimens. Journal of
Applied Physics, 72(8), 3305-3316.

\bibitem {ciava2021}Ciavarella, M. (2021). An upper bound for viscoelastic
pull-off of a sphere with a Maugis-Dugdale model. The Journal of Adhesion, 1-14.

\bibitem {AffVio2022}Afferrante, L., \& Violano, G. (2022). On the effective
surface energy in viscoelastic Hertzian contacts. Journal of the Mechanics and
Physics of Solids, 158, 104669.

\bibitem {VioAff2022size}Violano, G., \& Afferrante, L. (2022). Size effects
in adhesive contacts of viscoelastic media. European Journal of
Mechanics-A/Solids, 104665.

\bibitem {VioAffRange}Violano, G., \& Afferrante, L. (2022). On the long and
short-range adhesive interactions in viscoelastic contacts. Tribology Letters,
70(3), 1-5.

\bibitem {MuserPer}M{\"u}ser, M. H., \& Persson, B. N. (2022). Crack and
pull-off dynamics of adhesive, viscoelastic solids. Europhysics Letters,
137(3), 36004.

\bibitem {TangHui}Tang, T., \& Hui, C. Y. (2005). Decohesion of a rigid punch
from an elastic layer: Transition from \textquotedblleft flaw
sensitive\textquotedblright\ to \textquotedblleft flaw
insensitive\textquotedblright\ regime. Journal of Polymer Science Part B:
Polymer Physics, 43(24), 3628-3637.

\bibitem {Feng}Feng, J. Q. (2000). Contact behavior of spherical elastic
particles: a computational study of particle adhesion and deformations.
Colloids and Surfaces A: Physicochemical and Engineering Aspects, 172(1-3), 175-198.

\bibitem {Chri}Christensen, R. (2012). Theory of viscoelasticity: an
introduction. Elsevier.

\bibitem {persson2017}Persson, B. N. J. (2017). Crack propagation in
finite-sized viscoelastic solids with application to adhesion. EPL
(Europhysics Letters), 119(1), 18002.

\bibitem {PapCia2020}Papangelo, A., \& Ciavarella, M. (2020). A numerical
study on roughness-induced adhesion enhancement in a sphere with an
axisymmetric sinusoidal waviness using Lennard--Jones interaction law.
Lubricants, 8(9), 90.

\bibitem {ciavacricrimcmeek}Ciavarella, M., Cricr\`{\i}, G., \& McMeeking, R.
(2021). A comparison of crack propagation theories in viscoelastic materials.
Theoretical and Applied Fracture Mechanics, 116, 103113.

\bibitem {Joh}Johnson, K. L., (1987). Contact mechanics. Cambridge university press.

\bibitem {Kendall1971}Kendall, K. (1971). The adhesion and surface energy of
elastic solids. Journal of Physics D: Applied Physics, 4(8), 1186.

\bibitem {ciava2022}Ciavarella, M. (2022). Viscoelastic short cracks
propagation. Engineering Fracture Mechanics, 264, 108276.

{
\bibitem {Vio2023}Violano, G., De Carolis, S., Palmieri, M. E., Carbone, G.,
Tricarico, L., Demelio, G. P., \& Afferrante, L. (2023). Crack propagation
in viscoelastic finite-sized solids: theory and experiments. In IOP
Conference Series: Materials Science and Engineering (Vol. 1275, No. 1, p.
012043). IOP Publishing.}

\end{thebibliography}
\end{document}